\documentclass{PoS}

\title{Rare $B$ decays at CMS}

\ShortTitle{Rare $B$ decays at CMS}

\author{\speaker{Niladribihari Sahoo}\thanks{On behalf of the CMS Collaboration.}\\
        School of Physical Sciences, NISER, Bhubaneswar, INDIA\\
        E-mail: \email{niladribihari.sahoo@cern.ch}}

\abstract{The Flavor Changing Neutral Current mediated decays $B \rightarrow \mu^{+}\mu^{-}$ and 
$B^{0} \rightarrow K^{*0}\mu^{+}\mu^{-}$ provide high sensitivity to new physics contributions. Sensitive observables 
include the branching fraction, the muon forward-backward asymmetry, the fraction of $K^{*0}$ longitudinal polarisation and 
the differential branching fraction. We report herein the recent results from CMS on these decay modes.}

\FullConference{9th International Workshop on the CKM Unitarity Triangle\\
		28  November - 3 December 2016\\
		Tata Institute for Fundamental Research (TIFR), Mumbai, India}

\begin{document}

\section{Introduction}
Heavy flavor physics is the study of high energy hadronic interactions among quark flavors.
The production and decay of heavy flavor play an important role in testing Quantum Chromodynamics (QCD) 
based calculations. 
%%Here we discuss the production and decay of $B$ mesons containing $b$ quarks. 
They provide ample experimental observables to study underlying physics enabling us to 
search for new phenomena beyond the standard model (SM) by comparing the experimental results 
with theoretical predictions. Any significant deviation of the measured observable from SM 
predictions would hint the presence of new physics (NP). In this paper, the heavy flavor results involving b-hadrons
obtained by the CMS~\cite{ref0} experiment, based on the data collected at 7/8/13 TeV center-of-mass energy, are presented.

\section{Dimuon spectrum with 13 TeV data}
\begin{figure}[!htbp]
  \begin{center}
     \includegraphics[width=.9\textwidth,height=10cm,keepaspectratio]{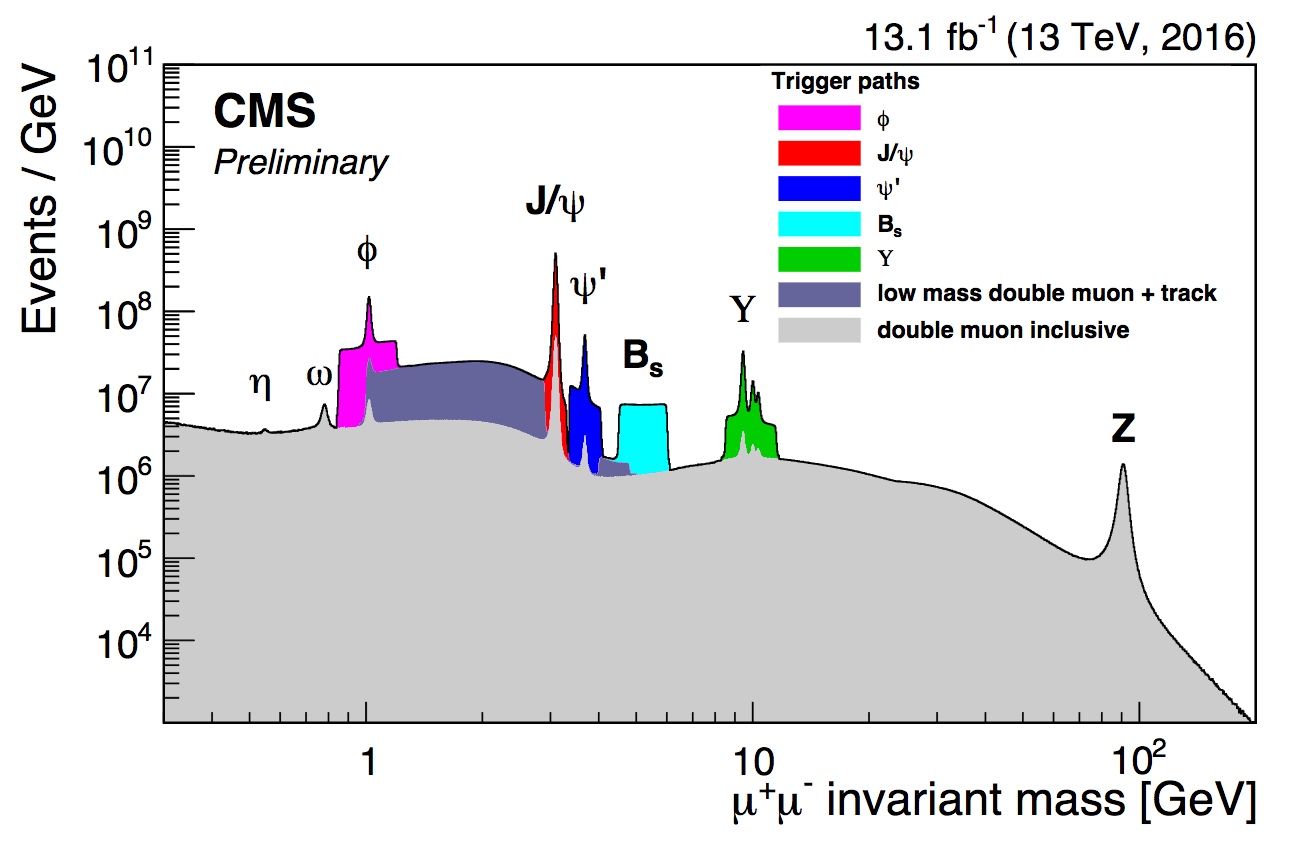}
     \caption{Dimuon mass distribution obtained with various dimuon triggers, during the 13 TeV data 
       taking in 2016. Different resonances are clearly visible in the distribution.}
     \label{fig1}
  \end{center}
\end{figure}

After the first long shut down of the LHC machine, the pp collisions have started again around middle
of 2015. The center-of-mass energy was 13 TeV corresponding to this LHC run. The CMS, ATLAS 
and LHCb experiments have already produced many results using 13 TeV data. CMS could reproduce 
the dimuon mass spectrum, as seen earlier with 7 and 8 TeV data samples. Figure~\ref{fig1} shows the dimuon 
invariant mass with 13.1 fb$^{-1}$ of data collected at 13 TeV during 2016~\cite{ref1}. The colored 
paths correspond to dedicated dimuon triggers with low $p_{T}$ thresholds, in specific mass windows, 
while the light gray continuous distribution represents events collected with a dimuon trigger with high $p_{T}$
threshold. One can clearly see contributions from different resonances such as $\phi$, J/$\psi$, $\psi^{'}$ and 
$\omega$ in the distribution.

\section{$B^{+}$ production cross-section with 13 TeV data}
\begin{figure}[!htbp]
  \begin{center}
     \includegraphics[width=.3\textwidth,height=10cm,keepaspectratio]{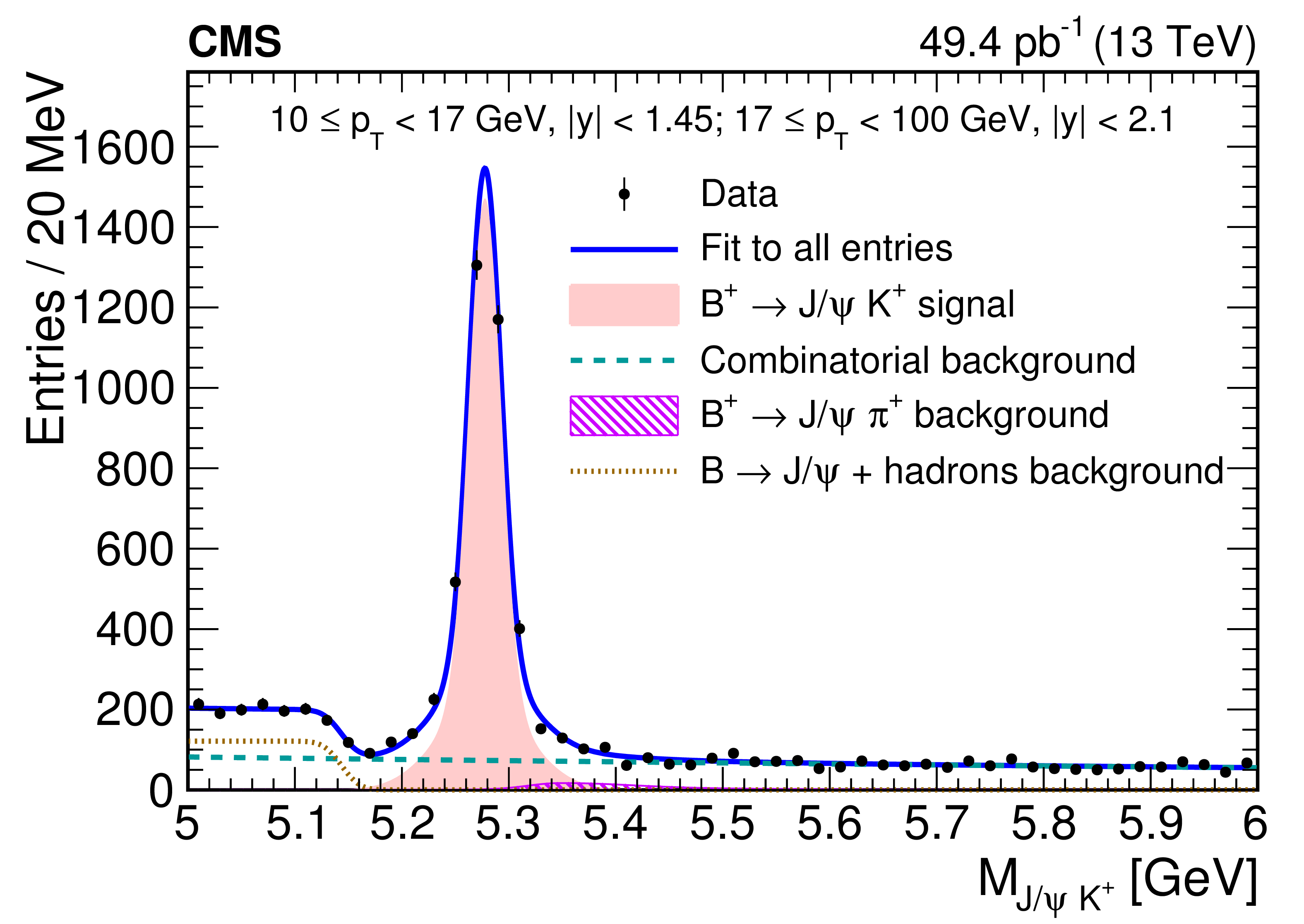}
     \includegraphics[width=.3\textwidth,height=10cm,keepaspectratio]{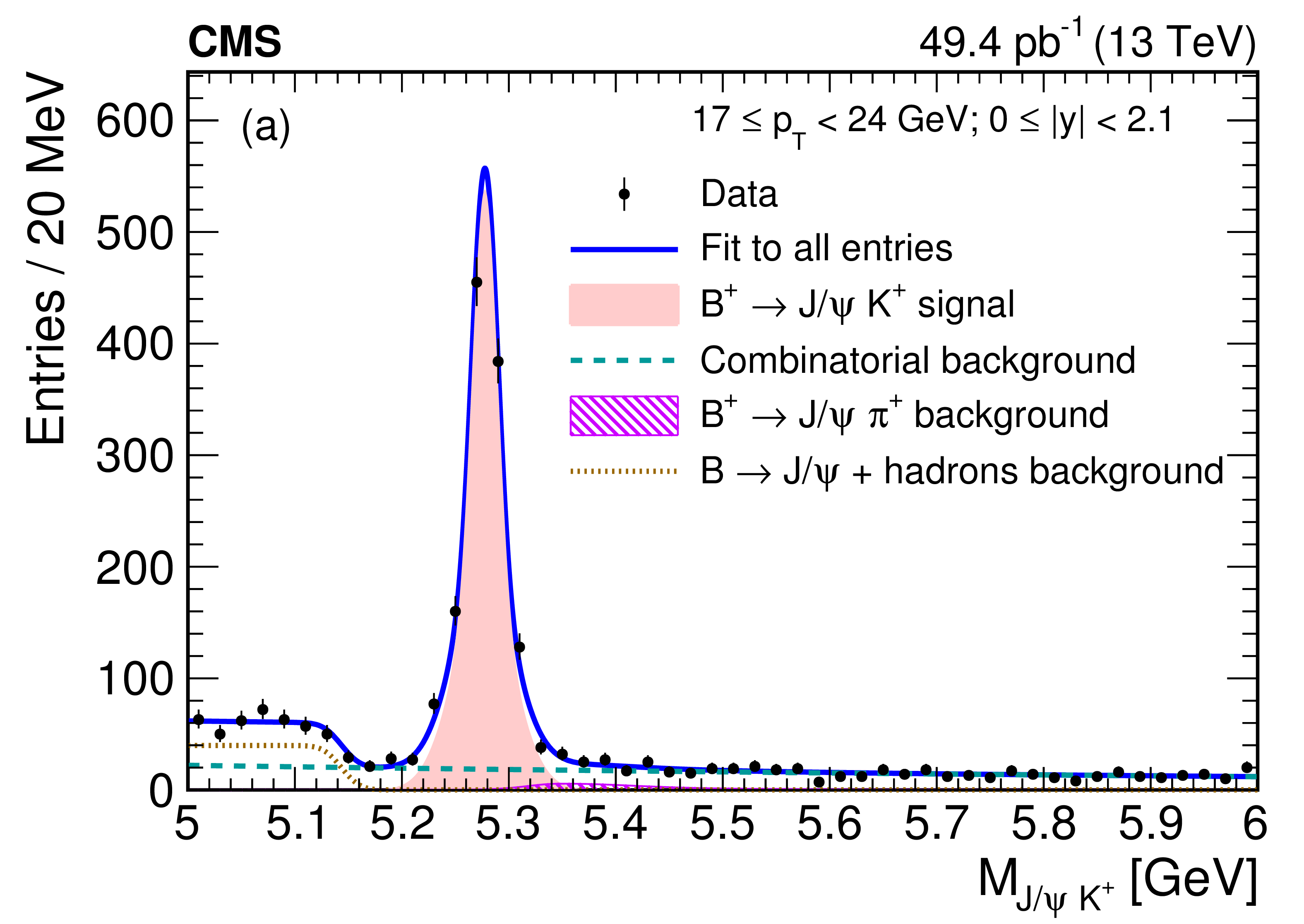}
     \includegraphics[width=.3\textwidth,height=10cm,keepaspectratio]{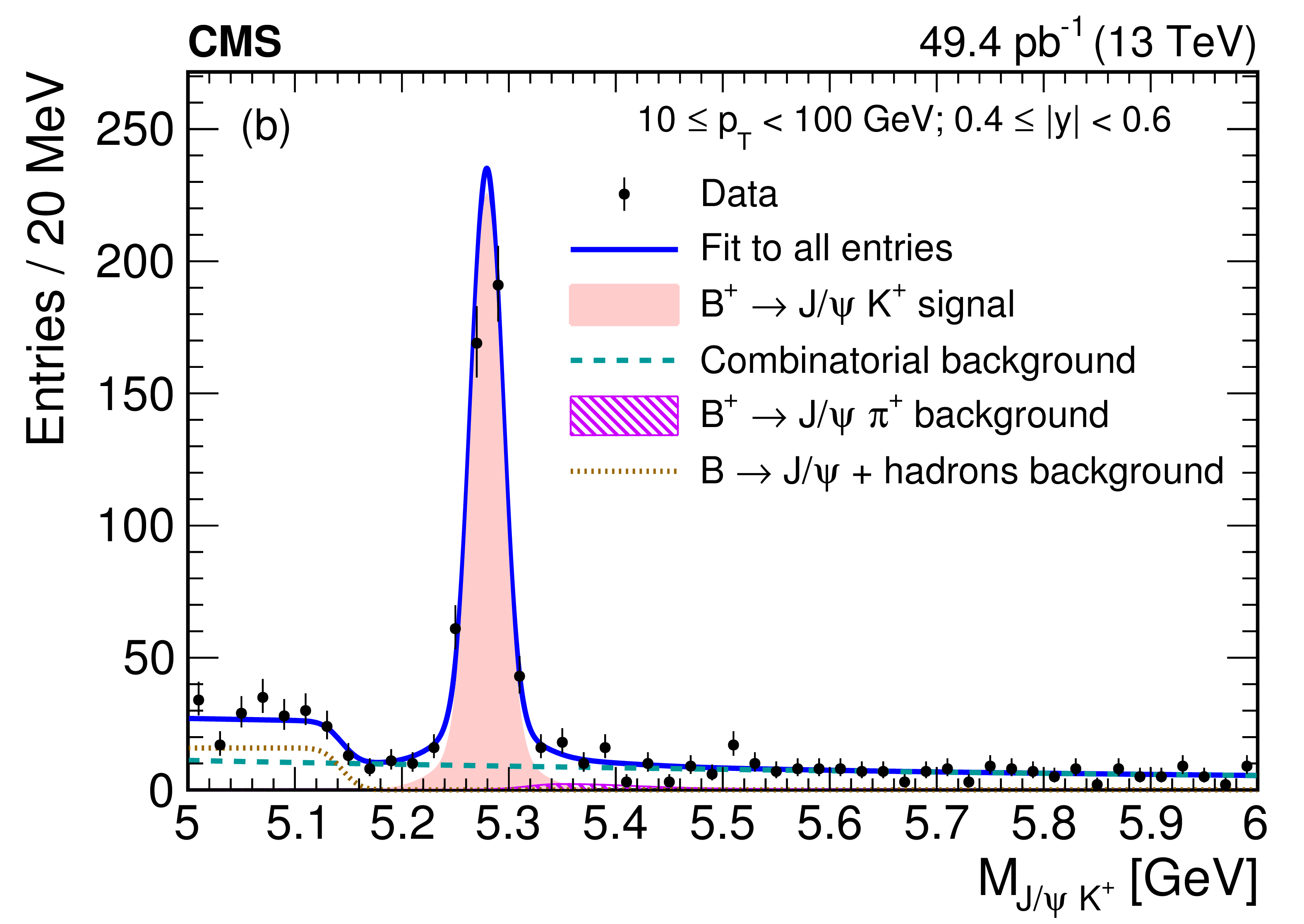}
     \caption{Invariant mass distribution of the $B^{+} \rightarrow J/\psi K^{+}$ candidates (left) 
       integrated in the phase-space region $10 < p^{B}_{T} < 17$~GeV with $|y^{B}| < 1.45$ and $17 < p^{B}_{T} < 100$~GeV 
       with $|y^{B}| < 2.1$, (middle) in the region $17 < p^{B}_{T} < 24$~GeV with $|y^{B}| < 2.1$, and (right) 
       $10 < p^{B}_{T} < 100$~GeV with $0.4 < |y^{B}| < 0.6$.}
     \label{fig2}
  \end{center}
\end{figure}

The measurements of $b$-hadron production cross sections provide essential information
to understand QCD. Such studies have been carried out by several collider experiments.
The recent studies of $b$-hadron production at the higher energies of the LHC Run 2 
provide a new important test of theoretical calculations. We present herein the
measurement of the $B^{+}$ differential cross-section in pp collisions at 13 TeV, as function
of transverse momentum ($p^{B}_{T}$) and rapidity ($y^{B}$). The result is based on a data sample
collected by the CMS experiment, corresponding to an integrated luminosity of 49.4 pb$^{-1}$, and uses 
the channel $pp \rightarrow B^{+}X \rightarrow J/\psi K^{+}X$ for selecting events where $J/\psi$ decays to 
a pair of muons. The muons are required to have at least one reconstructed segment in the muon stations 
that matches the extrapolated position of a track reconstructed in the silicon tracker satisfying $p_{T} > 4.2$~GeV 
and $|\eta| < 2.1$ and to have good quality in the fit to a track. The J/$\psi$ candidate must have an invariant mass 
within $\pm 150$~MeV of the nominal J/$\psi$ mass~\cite{pdg16}. Each J/$\psi$ candidate must have $p_{T} > 8$~GeV 
and the $\chi^{2}$ probability of the dimuon vertex fit is required to be larger than 10\%. Both muons must be either 
within $|\eta| < 1.6$ or one of them must have $p_{T} > 11$~GeV. $B^{+}$ candidates are reconstructed by combining 
a J/$\psi$ candidate with a charged track of $p_{T} > 1$~GeV. The track is assumed to be a kaon and the track-fit 
$\chi^{2}$ must be less than five times the number of degrees of freedom. To reduce the peaking and 
non-peaking backgrounds, many kinematic and topological cuts have
been applied. Finally, the signal yield is extracted with an extended unbinned maximum likelihood fit to the invariant mass of
the $B^{+}$ candidates in each of the $p^{B}_{T}$ and $y^{B}$ bins. Figure~\ref{fig2} shows the invariant mass 
distribution of all the $B^{+}$ candidates included in the analysis together with the corresponding fit results~\cite{ref2}.

\begin{figure}[!htbp]
  \begin{center}
     \includegraphics[width=.45\textwidth,height=10cm,keepaspectratio]{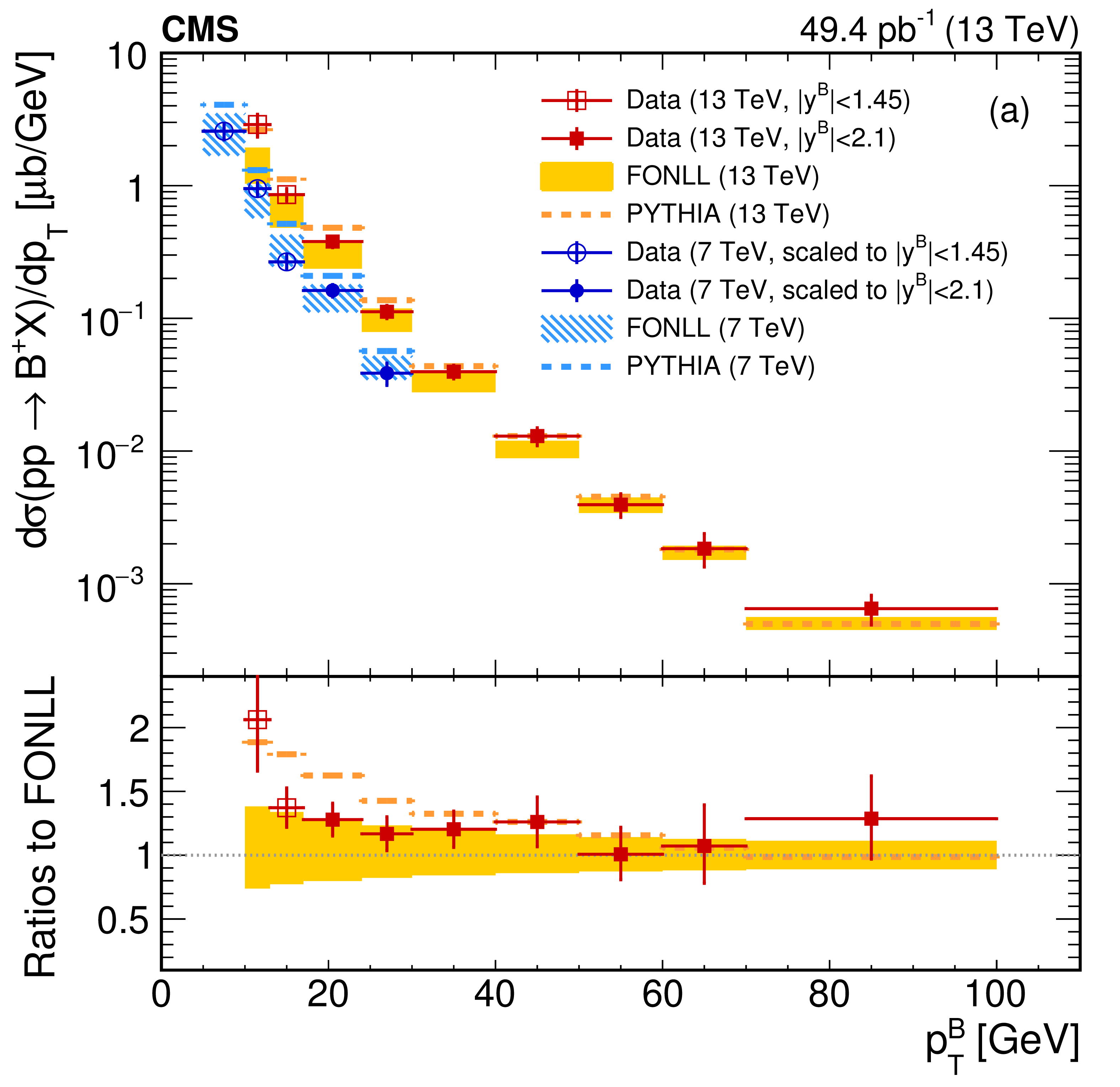}
     \includegraphics[width=.45\textwidth,height=10cm,keepaspectratio]{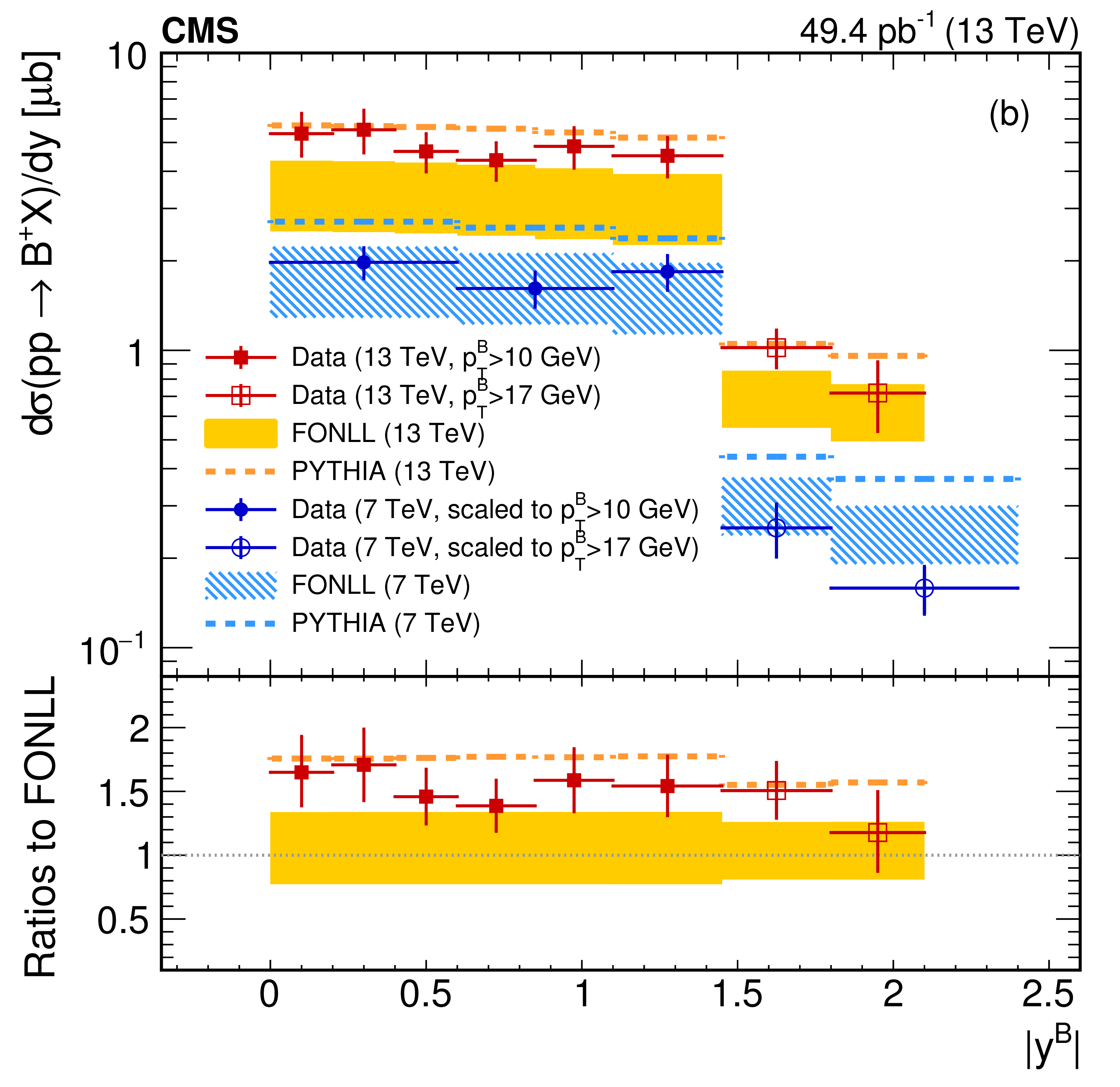}
     \caption{Differential cross section as a function of $p^{B}_{T}$ (left) and $y^{B}$ (right).}
     \label{fig3}
  \end{center}
\end{figure}

The differential cross sections as a function of $p^{B}_{T}$, integrated within $y^{B} < 1.45$ or for $y^{B} < 2.1$, and as a 
function of $y^{B}$, with $p^{B}_{T} > 10$~GeV or $p^{B}_{T} > 17$~GeV, are shown in Figure~\ref{fig3} where they 
are compared to FONLL (shaded boxes) and PYTHIA (dashed lines) calculations. The 7 TeV results are also
displayed on the plots for completeness. The bottom panels of Figure~\ref{fig3} display the data over FONLL 
cross-section ratios; the PYTHIA over FONLL ratios are also shown as dashed lines. The measured values show 
a reasonable agreement, both in terms of shape and normalization with FONLL calculations and with the results 
obtained with PYTHIA event generator.

\section{Study of {\boldmath $B^{0}_{s} \rightarrow \mu^{+}\mu^{-}$}}
In the SM, tree level diagrams do not contribute
to Flavor Changing Neutral Current (FCNC) mediated decays. Such decays mainly proceed
through higher-order loop diagrams, opening up the possibility for contributions from
non-SM particles. Within the SM, the rare FCNC decays $B \rightarrow \mu^{+}\mu^{-}$ have small 
branching fractions, i.e, $\mathcal{B}(B_{s}^{0} \rightarrow \mu^{+}\mu^{-}) = (3.66 \pm 0.23) \times 10^{-9}$ 
and $\mathcal{B}(B^{0} \rightarrow \mu^{+}\mu^{-}) = (1.06 \pm 0.09) \times 10^{-10}$~\cite{smexp}. 
%%%Charge conjugation is implied throughout this Letter. 
Several extensions of the SM, such as supersymmetric
models with non-universal Higgs boson masses, models containing leptoquarks,
and the minimal supersymmetric standard model with large tan$\beta$, predict enhancements
to the branching fractions for these rare decays. The decay rates can also be suppressed
for specific choices of model parameters. Over the past 30 years, a significant progress in
the search sensitivity has been made with exclusion limits on the branching fractions improving by
five orders of magnitude.

A search for the $B\rightarrow \mu^{+}\mu^{-}$ signal, where $B$ denotes $B^{0}_{s}$ or $B^{0}$, is performed by 
the CMS experiment in the dimuon invariant mass regions around the nominal $B^{0}_{s}$ or $B^{0}$ mass. To avoid possible
biases, the signal region $5.20 < m_{\mu\mu} < 5.45$~GeV was kept blind until all selection criteria
were established. The $B \rightarrow \mu^{+}\mu^{-}$ candidates are reconstructed from two oppositely charged muons 
that are required to have $p_{T} > 4$~GeV and be consistent in direction and momentum with the muons that triggered the event. A boosted 
decision tree (BDT) constructed within the TMVA framework~\cite{tmva} is trained to further separate genuine muons from those 
arising from misidentified charged hadrons. The variables used in the BDT can be divided into four classes: basic kinematic
quantities, silicon-tracker fit information, combined silicon and muon track fit information, and muon detector information.  
The BDT is trained on MC simulation samples of $B$ meson decays to kaons and muons. Compared to the ``tight'' muons, the BDT 
working point used to select muons for this analysis reduces the hadron-to-muon misidentification probability by 50\% while 
retaining 90\% of true muons. The probability to misidentify a charged hadron as a muon because of decay in flight or 
detector punch-through is measured in data from samples of well-identified pions, kaons and protons. This probability ranges
from $(0.5 - 1.3) \times 10^{-3}$, $(0.8 - 2.2) \times 10^{-3}$, and $(0.4 - 1.5) \times 10^{-3}$ for pions, kaons, 
and protons, respectively, depending on whether the particle is in the barrel or endcap, the running period, and the momentum. 
Each of these probabilities is assigned an uncertainty of 50\%, based on differences between data and MC simulation.

\begin{figure}[!htbp]
  \begin{center}
    \includegraphics[width=.45\textwidth,height=10cm,keepaspectratio]{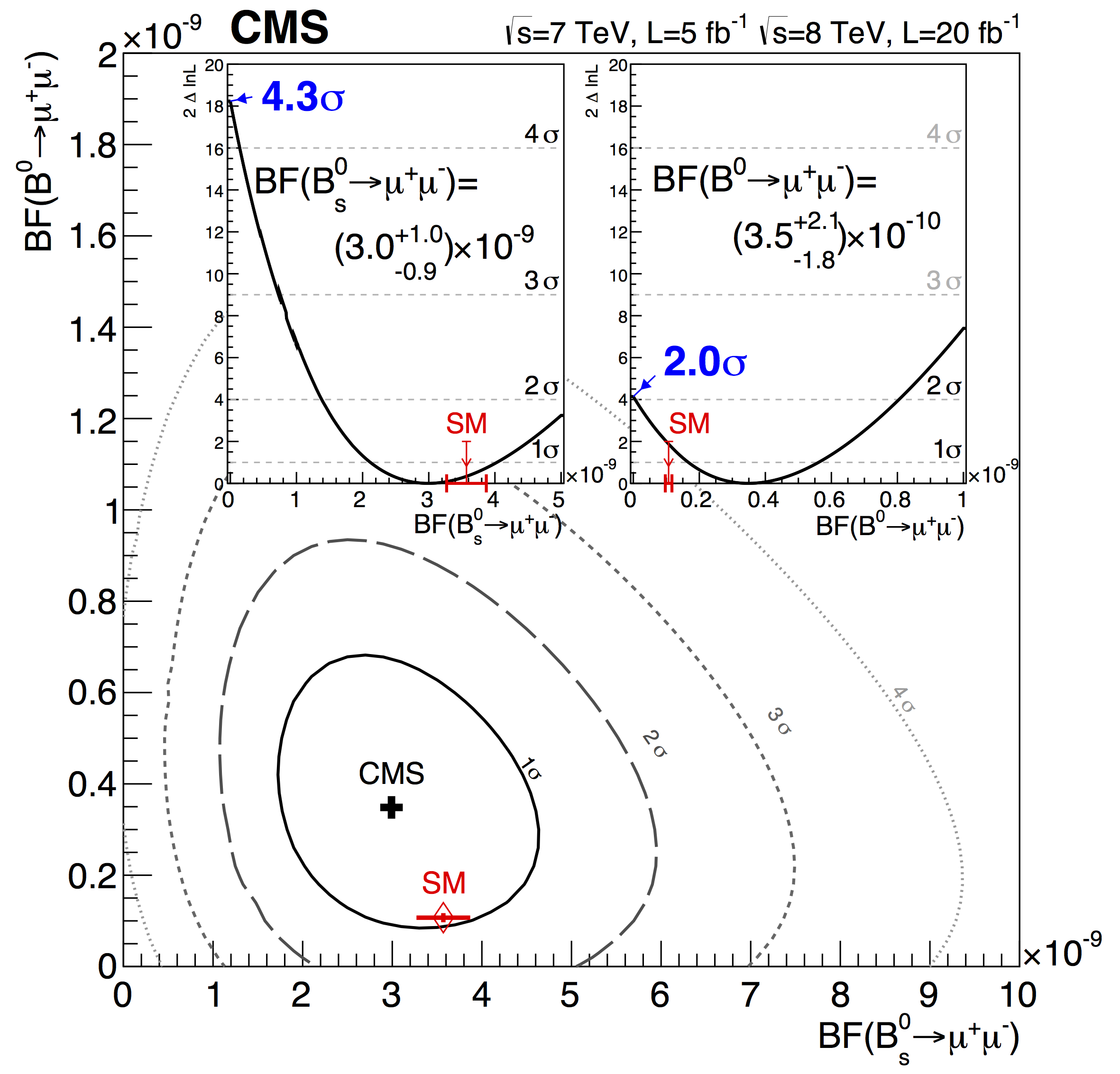}
    \includegraphics[width=.45\textwidth,height=10cm,keepaspectratio]{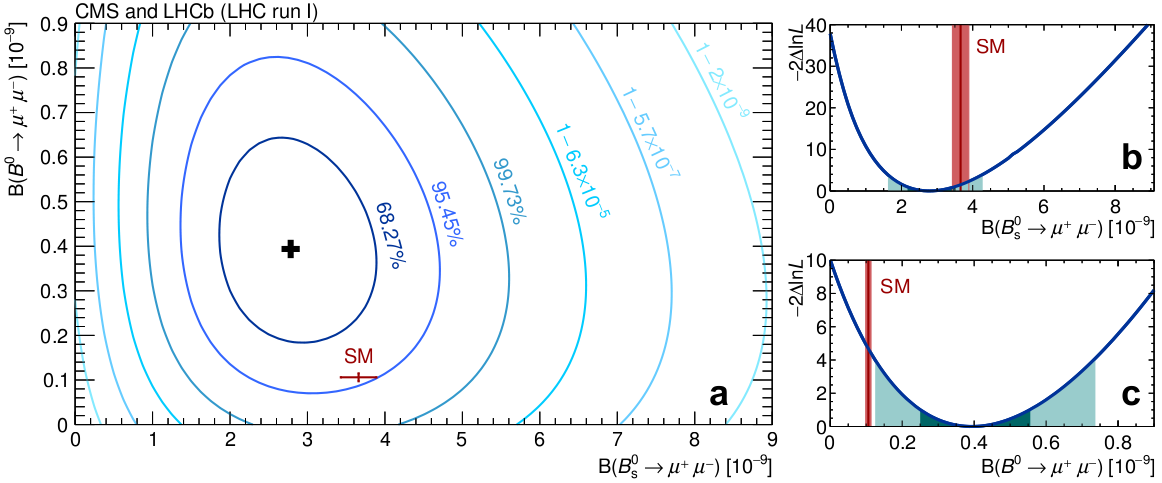}
     \caption{(Left) CMS result for the scan of joint likelihood ratio for $B^{0}_{s} \rightarrow \mu^{+}\mu^{-}$ and 
       $B^{0} \rightarrow \mu^{+}\mu^{-}$. As insets, the likelihood ratio scan is shown for each of the branching fractions when 
       the other is profiled together with other nuisance parameters; the significance at which the background-only 
       hypothesis is rejected is also shown. (Right) CMS and LHCb combined result showing the likelihood contours 
       in $\mathcal{B}(B^{0} \rightarrow \mu^{+}\mu^{-})$ versus $\mathcal{B}(B^{0}_{s} \rightarrow \mu^{+}\mu^{-})$ plane. 
       Also shown are the variations of the test statistics for $\mathcal{B}(B^{0} \rightarrow \mu^{+}\mu^{-})$ and 
       $\mathcal{B}(B^{0}_{s} \rightarrow \mu^{+}\mu^{-})$.}
     \label{fig4}
  \end{center}
\end{figure}

An unbinned maximum-likelihood fit to the $m_{\mu\mu}$ distribution is used to extract the
signal and background yields. Events in the signal window can result from genuine signal, 
combinatorial background, background from semileptonic $b$-hadron decays, and peaking background.  
The probability density functions for the signal, semileptonic and peaking backgrounds are obtained 
from fits to MC simulation. The dimuon mass distributions for the four channels (barrel and endcap in 7 and 8~TeV data), 
further divided into categories corresponding to different bins in the BDT output, are fitted simultaneously. 
No significant excess is observed for $B^{0} \rightarrow \mu^{+}\mu^{-}$ and an upper limit of 
$\mathcal{B}(B^{0} \rightarrow \mu^{+}\mu^{-}) < 1.1 \times 10^{-9}$ is set at 95\% confidence level. 
For $B^{0}_{s} \rightarrow \mu^{+}\mu^{-}$, an excess of events with significance of 4.3 standard deviations is 
observed, and a branching fraction of $\mathcal{B}(B^{0}_{s} \rightarrow \mu^{+}\mu^{-}) = 3.0^{+1.0}_{-0.9} \times 10^{-9}$
is determined, in agreement with the SM expectation~\cite{ref3}. The CMS and LHCb collaborations reported the combined 
result~\cite{ref4} for $B \rightarrow \mu^{+}\mu^{-}$ to exploit fully the statistical power of the data while accounting
 for the main correlations between them. The results are shown in Figure~\ref{fig4}.

\section{Angular analysis of the decay {\boldmath{$B^{0} \rightarrow K^{*0}\mu^{+}\mu^{-}$}}}
The FCNC decay $B^{0} \rightarrow K^{*0}\mu^{+}\mu^{-}$ provides many observables to search for new phenomena beyond the SM. 
The most notable ones are the forward-backward asymmetry 
of the muons, $A_{FB}$, the longitudinal polarization fraction of the $K^{*0}$, $F_{L}$, and the differential branching 
fraction, $\frac{d\mathcal{B}}{dq^{2}}$. To better decipher NP effects, 
these quantities can be measured as a function of the dimuon invariant mass squared ($q^{2}$). 

The reconstruction of decay candidates 
requires two muons of opposite charge and two oppositely charged hadrons. The muons are required to match those that triggered 
the event readout, and also to pass general muon identification requirements~\cite{muonid}. The hadron tracks, required to fail the muon 
identification criteria, must have $p_{T} > 0.8$~GeV and have a distance of closest approach to the beamspot in the 
transverse plane greater than twice the sum in quadrature of the distance uncertainty and the beamspot transverse size. 
The two hadrons with an invariant mass within 90~MeV of the nominal $K^{*0}$ mass are selected for further consideration. 
The $B^{0}$ candidates are obtained by fitting the four charged tracks to a common vertex, which leads to an improvement in  
the resolution of the track parameters. Other selection requirements are also applied to further purify the selected sample. 

\begin{figure}[!htbp]
  \begin{center}
     \includegraphics[width=.45\textwidth,height=10cm,keepaspectratio]{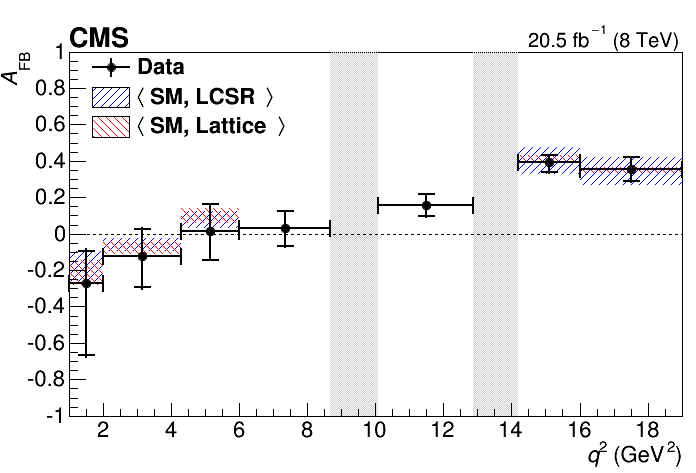}
     \includegraphics[width=.45\textwidth,height=10cm,keepaspectratio]{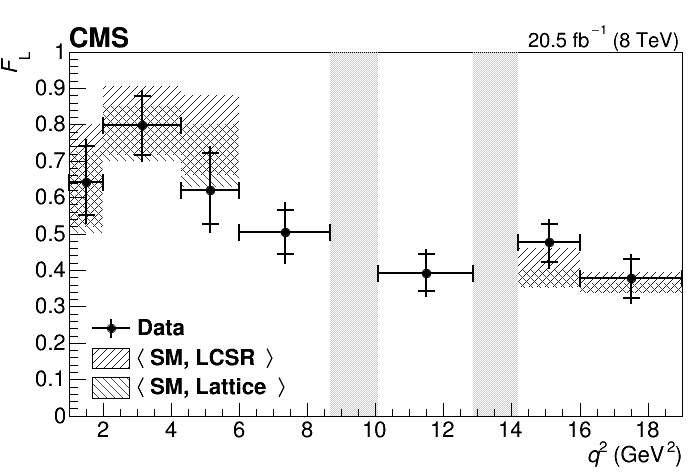}
     \includegraphics[width=.45\textwidth,height=10cm,keepaspectratio]{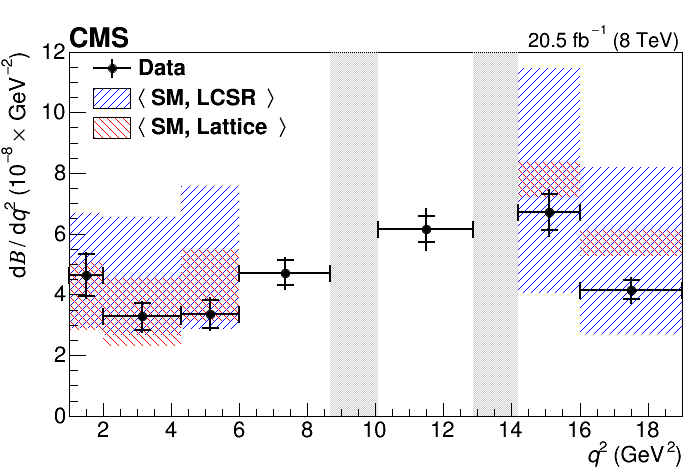}
     \caption{Measured values of $A_{FB}$, $F_{L}$ and $\frac{d\mathcal{B}}{dq^{2}}$ 
       versus $q^2$ for $B^{0} \rightarrow K^{*0}\mu^{+}\mu^{-}$. The statistical
       uncertainty is shown by the inner vertical bars, while the outer vertical bars give the total uncertainty. 
       The horizontal bars show the bin widths. The vertical shaded regions correspond to the
       J/$\psi$ and $\psi'$ resonances. The other shaded regions show the two SM predictions after 
       rate averaging across the $q^2$ bins to provide a direct comparison to the
       data. As theory predictions are not available for the J/$\psi$ and $\psi'$ resonance regions, we have not 
       considered them in our study.}
     \label{fig5}
  \end{center}
\end{figure}

For each $q^{2}$ bin, the observables of interest are extracted from an unbinned extended
maximum likelihood fit to three variables: the $B^{0}$ invariant mass and the two angular variables $\theta_{K}$
and $\theta_{L}$. Here, $\theta_{K}$ is the angle between the kaon momentum and the direction opposite to the $B^{0}$ ($\bar{B^0}$) in the 
$K^{*0}$ ($\bar{K^{*0}}$) rest frame, and $\theta_{L}$ is the angle between the positive (negative) muon momentum and the direction 
opposite to the $B^0$ ($\bar{B^0}$) in the dimuon rest frame. We find the results (presented in Figure~\ref{fig5}) to be 
consistent with SM predictions and previous measurements~\cite{ref5}. 

\section{Conclusions}
The results presented here show excellent prospects for rare $B$ decays at CMS. The Run1 and Run2 results are in 
good agreement with the theoretical predictions as well as with other experiments. CMS is collecting data at higher 
center-of-mass energy in Run2. At the end of Run2, the collected statistics would be enough in various cases to reach the 
level of NP prediction.

\end{document}